\documentclass[aps,prx,twocolumn,showpacs,superscriptaddress,amsmath,amssymb]{revtex4-1}

\usepackage{graphicx}
\usepackage[colorlinks=true,linkcolor=blue,citecolor=blue,urlcolor=blue]{hyperref}
\usepackage{bbold}
\usepackage{gensymb}

\def \a {\alpha}

\def \d {\delta}
\def \D {\Delta}

\def \ve {\varepsilon}

\def \g {\gamma}

\def \k {\kappa}

\def \L {\Lambda}
\def \o {\omega}

\def \t {\theta}

\def \s {\sigma}
\def \z {\zeta}
\def \dag {\dagger}

\def \p {\partial}
\def \dge {\degree}

\def \apx {\approx}

\def \til {\tilde}

\def \dag {\dagger}

\newcommand{\intv}[1]{\int_{\mbf #1}}

\newcommand{\sumv}[1]{\sum_{\mbf #1}}

\newcommand{\sumvdp}[1]{\sum_{\mbf #1\mbf #1'}}

\def \uar {\uparrow}
\def \dar {\downarrow}

\def \rar {\rightarrow}

\def \la {\langle}
\def \ra {\rangle}
\def \fr {\frac}
\def \lf {\left}
\def \ri {\right}

\newcommand{\ket}[1]{|#1\ra}

\newcommand{\epvl}[1]{\la#1\ra}

\def \Tr {\mathrm{Tr}}
\def \bece {\begin{center}}
\def \ence {\end{center}}
\def \beeq {\begin{equation}}
\def \eneq {\end{equation}}
\def \beal {\begin{aligned}}
\def \enal {\end{aligned}}
\def \bega {\begin{gathered}}
\def \enga {\end{gathered}}
\def \benu {\begin{enumerate}}
\def \ennu {\end{enumerate}}
\def \beit {\begin{itemize}}
\def \enit {\end{itemize}}
\def \bede {\begin{description}}
\def \ende {\end{description}}
\def \betb {\begin{tabular}}
\def \entb {\end{tabular}}
\def \bear {\begin{array}}
\def \enar {\end{array}}

\def \mbf {\mathbf}
\def \mbb {\mathbb}
\def \mrm {\mathrm}

\def \mca {\mathcal}

\def \bsb{\boldsymbol}

\begin{document}


\title{Kohn-Luttinger superconductivity on two orbital honeycomb lattice}

\author{Yu-Ping Lin}
\affiliation{Department of Physics, University of Colorado, Boulder, Colorado 80309, USA}
\author{Rahul M. Nandkishore}
\affiliation{Department of Physics, University of Colorado, Boulder, Colorado 80309, USA}
\affiliation{Center for Theory of Quantum Matter, University of Colorado, Boulder, Colorado 80309, USA}

\date{\today}

\begin{abstract}
Motivated by experiments on twisted bilayer graphene, we study the emergence of superconductivity from {\it weak} repulsive interactions in the Hubbard model on a honeycomb lattice, with both spin and orbital degeneracies, and with the filling treated as a tunable control parameter. The attraction is generated through the Kohn-Luttinger mechanism. We find, similar to old studies of single layer graphene, that the leading superconducting instability is in a $d$-wave pairing channel close to Van Hove filling, and is in an $f$-wave pairing channel away from Van Hove filling. The $d$-wave pairing instability further has a twelve-fold degeneracy while the $f$-wave pairing instability has a ten-fold degeneracy. We analyze the symmetry breaking perturbations to this model. Combining this with a Ginzburg-Landau analysis, we conclude that close to Van Hove filling, a spin singlet $d+id$ pairing state should form (consistent with several other investigations of twisted bilayer graphene), whereas away from Van Hove filling we propose an unusual spin and orbital singlet $f$-wave pairing state.
\end{abstract}

\maketitle

\section{Introduction}

Recent experiments observe superconductivity \cite{cao18nsc} proximate to an insulating state \cite{cao18nmi} in twisted bilayer graphene with magic twist angle $\t\apx1.05\dge$. Remarkably, the superconductivity exhibits a relatively high critical temperature $T_c$ with a small Fermi surface. To be explicit, the temperature ratio $T_c/T_F$, where $T_F$ is the Fermi temperature, is close but higher than most of the currently known high temperature superconductors. The carrier density-temperature phase diagram demonstrates two superconducting domes on both sides of the insulating phase. These features, which resemble those observed in high $T_c$ materials, have triggered an explosion of interest in the twisted bilayer graphene systems. Given the similarity of the observed phase diagram to that of high $T_c$ materials, it has been widely assumed that the pairing is mediated by repulsive electron-electron interactions, as in the high $T_c$ materials. We also make this assumption, although we cannot exclude the possibility of phonon mediated pairing.

The theoretical analysis of the twisted bilayer graphene system can be divided into two distinct parts: the development of a suitable model Hamiltonian capturing the key features of the problem, and the analysis thereof. It is known that the system exhibits a Moir\'e pattern \cite{trambly10nl,suarez10prb,bistritzer11pnas,santos12prb,nam17prb} at small twist angles, where a superlattice with extremely large unit cells emerges. The corresponding low energy regime manifests four nearly flat minibands and a relatively large gap from the other bands. A number of works have proceeded from here to derive an effective low energy theory \cite{xu18prl,yuan18prb,po18prx,kang18prx,koshino18prx,zou18prb}, for example, by deriving the symmetry allowed maximally localized Wannier orbitals. We will make use here of the model Hamiltonian from Ref.~\onlinecite{yuan18prb,koshino18prx}, which takes the form of a Hubbard model on the honeycomb superlattice with a two-fold orbital degeneracy in addition to spin degeneracy.

\begin{figure}[b]
\centering
\includegraphics[scale = 1]{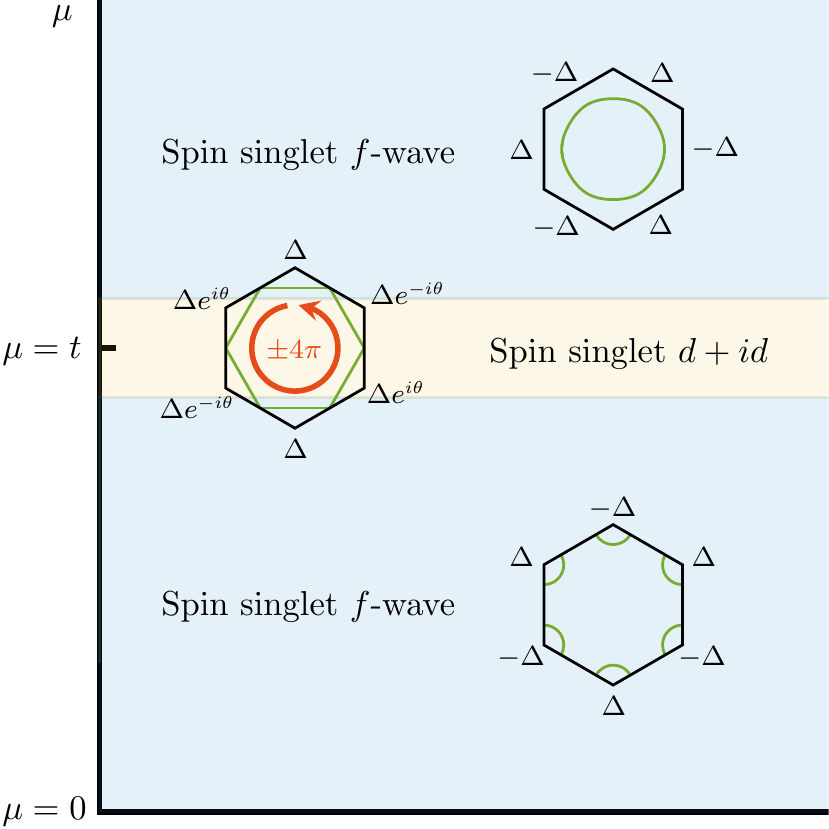}
\caption{\label{fig:phase} Phase diagram of superconductivities in the two orbital honeycomb lattice model. The black outer hexagons are the Brillouin zone boundaries of honeycomb lattice, while the inner green curves represent the Fermi surfaces. Near the Van Hove filling, a $d+id$ chiral superconductivity is dominant, within the Kohn-Luttinger approach. The phase of order parameter winds $\pm4\pi$ around the Fermi surface, where the angle $\t=\pm2\pi/3$ is defined. In the other regimes, a $f$-wave superconductivity is favored. Both of the dominant superconductivities manifest spin singlet pairing.}
\end{figure}

The question then arises as to how this model should be analyzed. Many investigations in this field have employed a {\it strong coupling} approach, assuming that the interactions are large compared to the bandwidth \cite{xu18prl, po18prx, dodaro18prb, rademaker18ax}. Here we take the opposite approach, and ask: What if the residual bandwidth is still large enough to exceed the interaction strength, such that the system is best described by a {\it weak coupling} approach? This approach can be motivated by considering, for example, twist angles slightly away from the magic angle. Our basic logic is to thoroughly explore the superconducting physics in the weak coupling regime, in the hope that this may provide insight into the experiments. We therefore analyze the onset of superconductivity in the model from Ref.~\onlinecite{yuan18prb,koshino18prx}, assuming weak coupling, at a range of doping levels. Since the experimental phase diagram looks similar to that for high $T_c$ materials, we assume that the superconductivity arises from a purely electron mechanism. Due to the large lattice constant in the Moir\'{e} superlattice, we further assume that the electron-electron repulsion can be well approximated by an onsite `Hubbard' repulsion. Finally, we assume that the pairing arises from the Kohn-Luttinger mechanism   \cite{kohn65prl,galitski03prb,gonzalez08prb,nandkishore12np,nandkishore14prb,gonzalez18ax}, which is the most generic method for obtaining pairing from repulsion at weak coupling. Of course, weak coupling superconductivity generically involves transition temperatures much lower than Fermi temperatures, whereas experimentally $T_c/T_F$ is not that small, so we do not expect this picture to be {\it quantitatively} accurate. Nevertheless, it may provide insight into what is going on.

We pay particular attention to the $\mrm{SU}(4)$ spin and orbital degeneracy, and to its lifting. The assumption of full $\mrm{SU}(4)$ symmetry greatly simplifies the analysis. We use the Fierz identity \cite{Vek, savary17prb,venderbos18prx,lin18prb} to decompose the interactions into pairing channels characterized by the $\mrm{SU(4)}$ irreducible representations. These pairings can be interpreted as combinations of singlets and triplets of orbital and spin pairings. Adopting a patch description of the Fermi surface, we perform the Kohn-Luttinger analysis and determine the potential superconducting states in a broad range of fillings (Fig.~\ref{fig:phase}). Near the Van Hove filling, we find dominance of $d$-wave superconductivity, whereas away from Van Hove filling, $f$-wave superconductivity dominates. These findings are similar to earlier investigations of Kohn-Luttinger superconductivity on the honeycomb lattice \cite{nandkishore14prb}. We then turn to the lifting of the SU(4) degeneracy. We show that an effective anti-Hund's coupling emerges from the renormalization of bare Hund's coupling. Combining this with a Ginzburg-Landau analysis, we conclude that close to Van Hove filling, a spin singlet $d+id$ pairing state should form (consistent with several other works). Away from Van Hove filling, we predict an unusual spin singlet and orbital singlet $f$-wave pairing state, which would be forbidden in the absence of orbital degeneracy by Fermi statistics.

We now briefly discuss how our work relates to the prior literature. Where the larger graphene literature goes, our work builds directly on old analyses of superconductivity in doped single layer graphene \cite{gonzalez08prb, nandkishore12np, nandkishore14prb}. The key difference from that older literature is in the incorporation of the orbital degeneracy (absent in single layer graphene), and the analysis of lifting the resulting enlarged degeneracy. Where the more recent twisted bilayer graphene literature is concerned, the closest parallels are to other works taking a weak coupling approach \cite{liu18prl,you18ax,guo18prb,gonzalez18ax,isobe18prx}. However, these works focus on the vicinity of Van Hove filling, whereas we consider a broader doping range. The methods employed are also different. In Ref.~\onlinecite{liu18prl, you18ax} the random phase approximation (RPA) is employed, whereas in Ref.~\onlinecite{isobe18prx} a parquet renormalization group analysis similar to Ref.~\onlinecite{nandkishore12np} is performed. In the immediate vicinity of Van Hove filling, the nesting of the Fermi surface gives rise to various competing instabilities, and this competition is difficult to resolve within RPA or Kohn-Luttinger type approaches. In contrast, the parquet renormalization group analysis \cite{isobe18prx} has the advantage of treating all instabilities on an equal footing, and predicting a leading instability. On the other hand, the Kohn-Luttinger analysis that we perform has the advantage of being extremely transparent, as well as easy to extend away from Van Hove filling.  It is in any case reassuring that RPA calculations, parquet renormalization group, and Kohn-Luttinger calculations all predict a spin singlet $d+id$ chiral superconductor near Van Hove filling (similar also to the old literature on single layer graphene \cite{nandkishore12np}). The Kohn-Luttinger analysis of Ref.~\onlinecite{gonzalez18ax} is an outlier, predicting $p$-wave pairing, in contrast to our analysis. We do not understand the discrepancy, but speculate that it comes from a different choice of model Hamiltonian. None of these works consider the pairing far from Van Hove filling, where we find an unusual spin singlet $f$-wave pairing state.

\section{Model}

We start with the analysis of an $\mrm{SU}(4)$ symmetric two orbital Hubbard model on the honeycomb lattice \cite{yuan18prb,koshino18prx}
\beeq
\label{eq:2ohcham}
H = -t\sum_{\epvl{ij}}\lf(c_i^\dag c_j+h.c.\ri)+\fr{U}{2}\sum_i\lf(c_i^\dag c_i\ri)^2-\mu\sum_ic_i^\dag c_i.
\eneq
Each vector $c_i=(c_{ix\uar},c_{ix\dar},c_{iy\uar},c_{iy\dar})^T$ describes the four onsite degrees of freedom composed of two orbitals $\tau=x,y$ and two spins $\s=\uar,\dar$. The corresponding Pauli matrices $\vec\tau=(\tau^0,\bsb\tau)$ and $\vec\s=(\s^0,\bsb\s)$, where $\tau^0=\s^0=\mbb1$, serve as convenient representations in later analysis. We choose the hopping constant $t>0$, the onsite repulsion $U>0$, and the chemical potential $\mu$ as real numbers, where the weak coupling condition $U\ll t$ is imposed. The lattice spacing is set as unity.

\subsection{Noninteracting theory}

The noninteracting theory manifests two bands with dispersion energies
\beeq
\pm\ve_{\mbf k} = \pm t\sqrt{1+4\cos\fr{3k_x}{2}\cos\fr{\sqrt3k_y}{2}+4\cos^2\fr{\sqrt3k_y}{2}}.
\eneq
Each band manifests four-fold degeneracy, corresponding to the $\mrm{SU}(4)$ symmetry of the model. We focus on the positive bands with $\mu>0$, while the analysis of negative bands is similar. The noninteracting Hamiltonian can be expressed as
\beeq
H_0 = \sumv{k}\xi_{\mbf k}c_{\mbf k}^\dag c_{\mbf k},
\eneq
where $c_{\mbf k}$ is redefined as the positive energy modes. The relative energy to the Fermi level is $\xi_{\mbf k}=\ve_{\mbf k}-\mu$.

The Brillouin zone is a hexagon with six corner Dirac points $(0,\pm4\pi/3\sqrt3)$ and $(\pm2\pi/3,\pm2\pi/3\sqrt3)$. Starting from the full filling $\mu=3t$, the decrease of filling manifests a deformation of Fermi surface from the center point to a hexagon at Van Hove filling $\mu=t$. The six corner Van Hove points $(\pm2\pi/3,0)$ and $(\pm\pi/3,\pm\pi/\sqrt3)$ exhibit logarithmically divergent densities of states, known as the Van Hove singularity. When the filling further decreases $0<\mu<t$, the Fermi surface splits into six distinct arcs, and shrinks into the Dirac points at the half filling $\mu=0$.

\subsection{Interaction and pairing channels}

Our main purpose is to probe the potential superconductivity induced by the Kohn-Luttinger renormalization \cite{kohn65prl}. Despite the constant repulsion in the bare theory, the renormalized interaction can acquire momentum dependence from the high order corrections. We take the interaction with general momentum dependence
\beeq
\label{eq:4fintham}
H_\mrm{int} = -\fr{1}{2}\sumvdp{k}V_{\mbf k-\mbf k'}(c_{\mbf k}^\dag c_{\mbf k'})(c_{-\mbf k}^\dag c_{-\mbf k'})
\eneq
as a starting point for the analysis of pairing channels. The zero momentum pairing and the minus sign are imposed as in conventional studies of superconductivity.

The Fierz identity is frequently utilized to derive the pairing channels from the four fermion interactions \cite{Vek,savary17prb,venderbos18prx,lin18prb}. A first attempt regards the constraint of Fermi statistics. We separate the four fermion part in Eq.~(\ref{eq:4fintham}) into two pairing channels
\beeq
(c_{\mbf k}^\dag c_{\mbf k'})(c_{-\mbf k}^\dag c_{-\mbf k'})
=
\fr{1}{4}\lf[(\vec P_{\mbf k}^s)^\dag\cdot\vec P_{\mbf k'}^s+(\vec P_{\mbf k}^a)^\dag\cdot\vec P_{\mbf k'}^a\ri].
\eneq
The pairing operators consist of the (anti)symmetric $\mrm{SU}(4)$ irreducible representations \cite{Vek,xu18prl}
\beeq
(\vec P_{\mbf k}^{s,a})^\dag=c_{\mbf k}^\dag\vec{\mca M}^{s,a}[\g(c_{-\mbf k}^\dag)^T],
\eneq
thereby feature the (anti)symmetric pairing of quantum numbers. Analogous to the time reversal operator $i\s^y$ for spin-$1/2$ systems, the unitary operator $\g = i(i\tau^y)(i\s^y)$ is defined with $\g^2=-1$. The $10(6)$ component vector $\vec{\mca M}^{s(a)}$ represents the (anti)symmetric $\mrm{SU}(4)$ irreducible representations normalized by $\Tr[\mca M^{s,a}_\z(\mca M^{s,a}_\z)^\dag]=4$, where $\z$ denotes the characteristic quantum number for a pairing channel.

The symmetric and antisymmetric pairing channels can be interpreted more clearly in the singlet-triplet representation. Consider the combinations of singlets and triplets between the orbital and spin pairings. While the symmetric pairings are composed of either two singlets or two triplets $\mca M^s_\z = \tau^0\s^0, \tau^i\s^j$, the antisymmetric pairings feature one singlet and one triplet $\mca M^a_\z = \tau^0\s^i, \tau^i\s^0$. With the aid from the orbital pairings, both spin singlet and triplet pairings can appear in the two pairing channels. This feature indicates that the model with two orbitals Eq.~(\ref{eq:2ohcham}) can potentially support pairing states that are unavailable in the usual systems.

The judgment of pairing channels also requires an analysis of the momentum dependent interaction $V_{\mbf k-\bf k'}$. Due to Fermi statistics, the interactions experienced by symmetric and antisymmetric quantum number pairings are
\beeq
V^{s,a}_{\mbf k-\mbf k'} = \fr{1}{2}\lf(V_{\mbf k-\mbf k'}\mp V_{\mbf k+\mbf k'}\ri),\quad
V^{s,a}_{\mbf k-\mbf k'} = \mp V^{s,a}_{\mbf k+\mbf k'}.
\eneq
The absence of $V^{a(s)}_{\mbf k-\mbf k'}$ for (anti)symmetric quantum number pairings can be confirmed by examining the cancellation between $\pm\mbf k'$ domains in the interaction Eq.~(\ref{eq:4fintham}).

From the analysis of compatibility with Fermi statistics, the Hamiltonians in the two pairing channels are determined
\beeq
\label{eq:chanmomham}
H^{s,a} =
\sumv{k}\xi_{\mbf k}c_{\mbf k}^\dag c_{\mbf k}
-\fr{1}{8}\sumvdp{k}V^{s,a}_{\mbf k-\mbf k'}(\vec P_{\mbf k}^{s,a})^\dag\cdot\vec P_{\mbf k'}^{s,a}
.
\eneq
In general, the interaction $V^{s,a}_{\mrm k-\mrm k'}$ is not diagonal in momentum space representation. This necessitates solving the related eigenvalue problem and evaluating the pairing channels with different momentum space configurations. We address this issue in the next subsection.

\subsection{Patch models}

\begin{figure}[t]
\centering
\includegraphics[scale = 1]{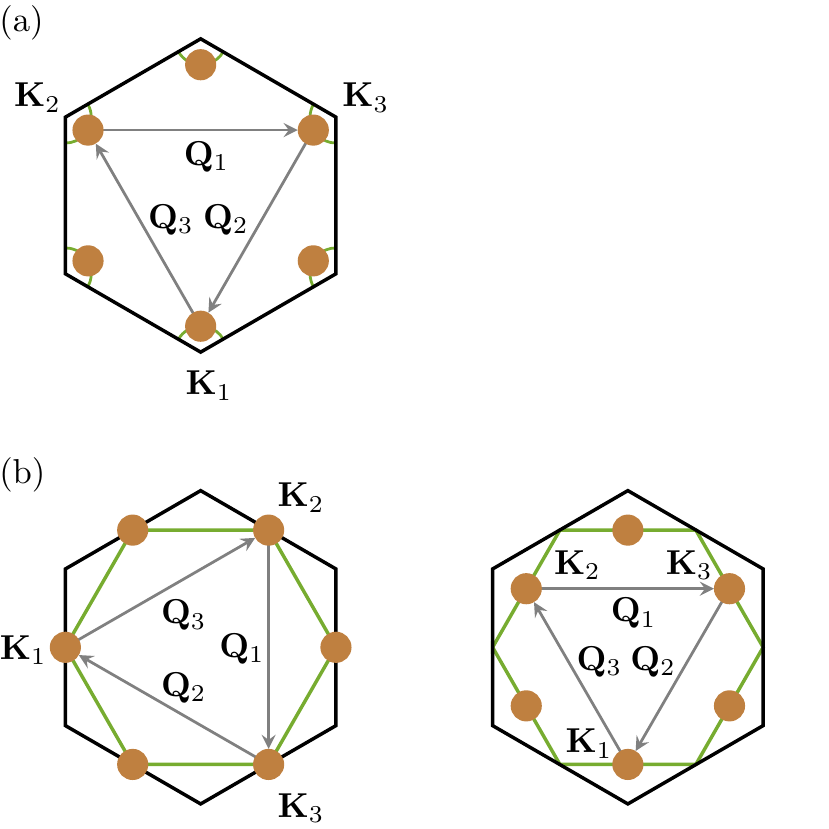}
\caption{\label{fig:patch} (a) Patch model II near the half filling. (b) Patch model I (left) and patch model II (right) near the Van Hove filling. The black outer hexagons indicate the Brillouin zone boundaries, while the green inner curves represent the Fermi surfaces. The patches are represented by the brown solid circles and are connected by the linking momenta $\mbf Q$'s.}
\end{figure}

In the low filling regime, the Fermi surface is dominated by some `hot spots'. These hot spots constitute a simplified patch model \cite{nandkishore12np,nandkishore14prb} for the Fermi surface. The choice of patch models depends strongly on the electron filling (Fig.~\ref{fig:patch}). Near Van Hove filling $\mu\apx t$, we conduct the analysis in two kinds of patch models. Patch model I exhibits the patches near the six Van Hove points, while in patch model II the patches sit near the six edge centers of Fermi surface. For the lower fillings $0<\mu<t$, only patch model II is adopted. The patches are now placed at the centers of the six Fermi arcs. In these patch models, the fermionic operators exhibit three components $c_\pm=(c_{\pm1},c_{\pm2},c_{\pm3})$ defined by the patch momenta $\mbf K_{\pm\a}$'s. In this convention, the momentum space summation becomes the summation over patch indices.

The pairing channels are determined by diagonalizing the interaction matrix $V^{s,a}$
\beeq
V^{s,a} = 
\lf(\bear{ccc}
V_0^{s,a}&V_Q^{s,a}&V_Q^{s,a}\\
V_Q^{s,a}&V_0^{s,a}&V_Q^{s,a}\\
V_Q^{s,a}&V_Q^{s,a}&V_0^{s,a}
\enar\ri).
\eneq
The diagonal elements $V^{s,a}_0=(V_{\mbf0}\mp V_{2\mbf K})/2$ are intrapatch interactions, while the offdiagonal ones $V^{s,a}_Q=(V_{\mbf Q}\mp V_{\mbf Q+2\mbf K})/2$ work between patches with linking momentum $\mbf Q$. After the diagonalization, the Hamiltonians in the pairing channels take the form
\beeq
\label{eq:chanham}
H^{s,a(i)} = 
\sum_{\k=\pm}\xi_\k c_\k^\dag c_\k
-\fr{1}{2}g^{s,a(i)}(\vec P^{s,a(i)})^\dag\cdot\vec P^{s,a(i)}
.
\eneq
The interactions $g^{s,a(i)}$'s are the eigenvalues of the interaction matrix $V^{s,a}$
\beeq
g^{s,a(0)} = V_0^{s,a}+2V_Q^{s,a},\quad
g^{s,a(1)} = g^{s,a(2)} = V_0^{s,a}-V_Q^{s,a}.
\eneq
Correspondingly, the pairing operators
\beeq
(\vec P^{s,a(i)})^\dag = c_+^\dag d^{(i)}\vec{\mca M}^{s,a}[\g(c_-^\dag)^T]
\eneq
are defined by the diagonal representations of orthonormal eigenstates $d^{(0)} = (1/\sqrt3)\mrm{diag}(1,1,1)$, $d^{(1)} = (1/\sqrt6)\mrm{diag}(2,-1,-1)$, and $d^{(2)} = (1/\sqrt2)\mrm{diag}(0,1,-1)$. Notice an extra factor of $4$ in the interaction due to the double counting of $\pm\a$ for the pairing operators.

\section{Superconductivity from weak electronic repulsion}

Having derived the pairing channels, the next task is to analyze the potential pairing instabilities that can arise in these channels. This seems impossible at first glance due to the bare repulsive interaction. However, as the high order corrections are taken into account, the repulsive interaction can be screened or even overscreened. The realization of attractive interaction and superconductivity therefore becomes possible. This mechanism is known as the Kohn-Luttinger renormalization \cite{kohn65prl}.

\subsection{Superconducting features of pairing channels}

Before embarking on the Kohn-Luttinger analysis, we inspect the potential superconducting features of the pairing channels \cite{cheng10prb,nandkishore14prb}. The standard analysis regards the gap function and the corresponding order parameter
\beeq
\label{eq:gapfunordpar}
\D^{s,a(i)}_{\a\eta\eta'} = d^{(i)}_{\a\a}\vec{\mca M}^{s,a}_{\eta\eta'}\cdot\vec\D^{s,a(i)},
\quad
\vec\D^{s,a(i)} = \fr{g^{s,a(i)}}{2}\epvl{\vec P^{s,a(i)}},
\eneq
where $\alpha$, $i$, and $\eta$ label the patches, eigenvectors, and quantum numbers, respectively. When the quantum number pairing is symmetric, the gap function is odd under inversion $\D^{s(i)}_\a = -\D^{s(i)}_{-\a}$. One $f$-wave and two degenerate $p$-wave pairing channels are identified. For the antisymmetric quantum number pairings, the even gap functions $\D^{a(i)}_\a = \D^{a(i)}_{-\a}$ indicate one $s$-wave and two degenerate $d$-wave pairing channels. We label these pairing channels by $l=s,p_1,p_2,d_1,d_2,f$ in the following context.

\subsection{Kohn-Luttinger renormalization}

\begin{figure}[b]
\centering
\includegraphics[scale = 1]{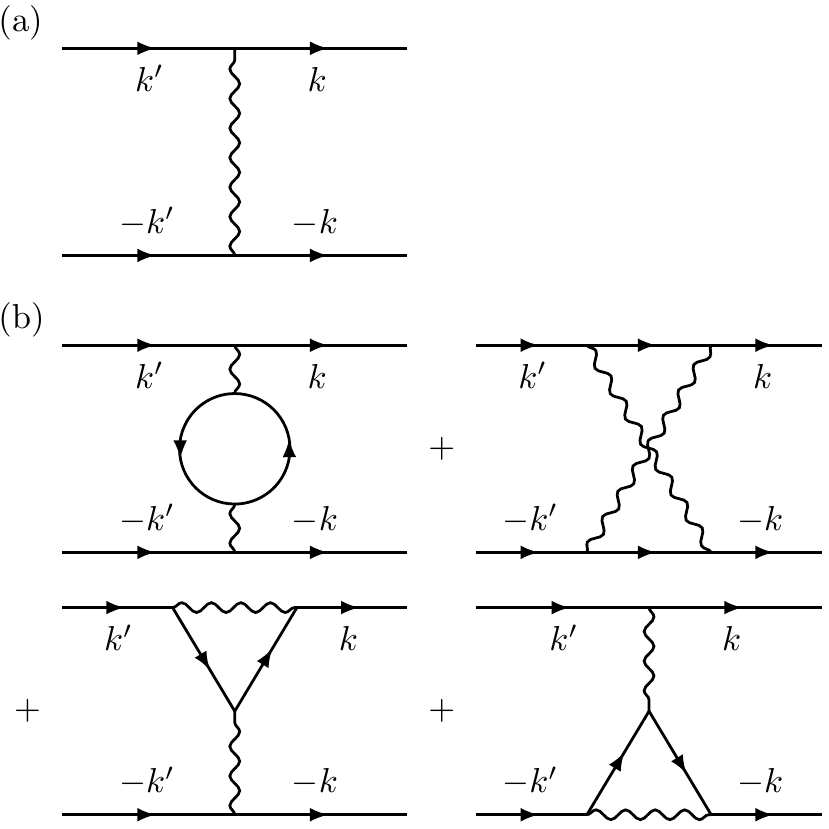}
\caption{\label{fig:diagram} (a) Particle-particle channel with bare repulsion. (b) The second order diagrams in the Kohn-Luttinger renormalization. The four diagrams are described by the polarization bubbles $4\Pi_{\mbf k-\mbf k'}$, $-\Pi_{\mbf k+\mbf k'}$, $-\Pi_{\mbf k-\mbf k'}$, and $-\Pi_{\mbf k-\mbf k'}$, respectively. A summation over these diagrams provides the correction to the interaction Eq.~(\ref{eq:khrint}).}
\end{figure}

In the original model, the only interaction is the constant onsite repulsion $V_{\mbf k-\mbf k'}=-U/2$. Only the $s$-wave pairing channel experiences a finite repulsion $g^s = -3U/2$. Under the Kohn-Luttinger renormalization, the interaction is corrected by the second order diagrams (Fig.~\ref{fig:diagram}). These diagrams are described by the static polarization bubbles $\Pi_{\mbf q}=-T\sum_\o\intv{p}G_{\mbf p\o}G_{(\mbf p+\mbf q)\o}$, where $G_{\mbf p\o}$ is the free fermionic propagator. The negative sign is introduced so that the polarization bubbles are positive semidefinite. With a summation over all diagrams, the correction to the interaction is determined
\beeq
\label{eq:khrint}
\d V_{\mbf k-\mbf k'} = \fr{U^2}{4}\lf(2\Pi_{\mbf k-\mbf k'}-\Pi_{\mbf k+\mbf k'}\ri).
\eneq
Notice that the first term is absent when there is only spin-$1/2$ degeneracy \cite{kohn65prl,galitski03prb,nandkishore14prb}, since the diagrams characterized by $\Pi_{\mbf k-\mbf k'}$ cancel out all together.

Calculating polarization bubbles is significantly simplified in the patch models. The domain of momentum integral in $\Pi_{\mbf q}$ reduces to the patch pairs with linking momentum $\mbf q$. We derive the renormalized interactions in the pairing channels
\beeq
\label{eq:intpairchan}
\beal
g^f &= \fr{3U^2}{8}\lf(\Pi_{\mbf0}-\Pi_{2\mbf K}+2\Pi_{\mbf Q}-2\Pi_{\mbf Q+2\mbf K}\ri),\\
g^p &= \fr{3U^2}{8}\lf(\Pi_{\mbf0}-\Pi_{2\mbf K}-\Pi_{\mbf Q}+\Pi_{\mbf Q+2\mbf K}\ri),\\
g^{s} &= -\fr{3U}{2}+\fr{U^2}{8}\lf(\Pi_{\mbf0}+\Pi_{2\mbf K}+2\Pi_{\mbf Q}+2\Pi_{\mbf Q+2\mbf K}\ri),\\
g^d &= \fr{U^2}{8}\lf(\Pi_{\mbf0}+\Pi_{2\mbf K}-\Pi_{\mbf Q}-\Pi_{\mbf Q+2\mbf K}\ri).
\enal\eneq
In $f$-wave and $p$-wave pairing channels, the only difference from single layer graphene is an extra factor of three in the corrections \cite{nandkishore14prb}. However, opposite signs arise in $s$-wave and $d$-wave pairing channels, necessitating a comprehensive reexamination.

For the single layer graphene, the dominant pairing channels at different fillings have been examined \cite{nandkishore14prb}. The analysis reveals $d$-wave superconductivity in the vicinity of Van Hove filling, while $f$-wave superconductivity takes over away from Van Hove filling. In the two orbital model adopted to describe twisted bilayer graphene, similar results are obtained. We briefly summarize our main results in the remaining part of this section. A comprehensive analysis is presented in Appendix \ref{sec:appkl}.

Away from Van Hove filling, all patch pairs provide similar contributions. The analysis can be executed by counting the available patch pairs in the polarization bubbles. We utilize patch model II near the half filling.  The evaluation of interactions Eq.~(\ref{eq:intpairchan}) reveals the dominance of $f$-wave superconductivity. Notice that the interaction is three times stronger than that in the single layer graphene. This enhancement is caused by the enlarged contribution from the internal fermion loop in two orbital model. Similar analysis also applies to the regimes near but finitely distant from Van Hove filling. With the examination of both patch models, the dominance of $f$-wave superconductivity is again confirmed.

The analysis becomes more complex in the vicinity of Van Hove filling. Due to the Van Hove singularity and the nesting of Fermi surface, the polarization bubbles acquire logarithmically divergent scalings as either $\ln(\L/T)$ or $\ln^2(\L/T)$. Here $\L$ is an ultraviolet cutoff determined by the patch size. In patch model I, an attractive correction with divergent scaling $\ln^2(\L/T)$ arises only in the $s$-wave pairing channel. However, the second order correction is not expected to overcome the bare repulsion in perturbation theory. On the other hand, patch model II exhibits attractive corrections in both $s$-wave and $d$-wave pairing channels, where the $s$-wave pairing channel again remains repulsive. This suggests the dominance of $d$-wave superconductivity in the vicinity of Van Hove filling, much as occurs for single layer graphene. Notice however that the attractive correction acquires a logarithmic divergence $\ln(\L/T)$, instead of the more divergent scaling $\ln^2(\L/T)$ arising in single layer graphene \cite{nandkishore14prb}.

Of course, the logarithmic divergences in polarization bubbles suggest the breakdown of perturbation theory. Furthermore, the existence of competing ordering tendencies near Van Hove filling implies that Kohn-Luttinger analysis alone is not trustworthy. The parquet renormalization group \cite{nandkishore12np,isobe18prx,lin18tbp} is the standard way to solve these problems. However, parquet analysis at Van Hove filling still yields the leading superconducting instability in the $d$-wave pairing channel \footnote{For the single layer graphene, the dominance of $d$-wave superconductivity over all instabilities is unambiguous \cite{nandkishore12np}. For the twisted bilayer graphene, however, there are also some competitions from density wave states and $p$-wave pairing channel \cite{isobe18prx, lin18tbp}. Since $p$-wave pairing does not dominate within our Kohn-Luttinger approach, we do not consider it further here.}. The conclusion of our Kohn-Luttinger analysis is therefore reinforced.

With the Kohn-Luttinger renormalization, the potential superconductivities at different fillings have been determined. We expect $d$-wave pairing close to Van Hove filling, and $f$-wave pairing away from Van Hove filling.

\section{Breakdown of degeneracies}

We have discussed the potential pairing instabilities in the model Eq.~(\ref{eq:2ohcham}). Large degeneracies are present due to various quantum number pairings and momentum space configurations. In practice, such degeneracies are likely lifted by symmetry breaking perturbations. Here we examine the splitting of degeneracies exhibited by the $d$-wave and $f$-wave superconductivity.

\subsection{$d$-wave superconductivity}

\subsubsection{Effective anti-Hund's coupling}

Since the $d$-wave pairing channels exhibit antisymmetric quantum number pairings, the breakdown of degeneracy can benefit from the onsite perturbations. In multiorbital systems, the spin configurations are usually determined by the Hund's coupling \cite{xu18prl,yuan18prb}
\beeq
\label{eq:hunds}
H_J = -J\sum_i\sum_{\tau\tau'}\mbf S_{i\tau}\cdot\mbf S_{i\tau'}.
\eneq
We consider the effect of perturbative Hund's coupling $J\ll U$ on the degenerate $d$-wave pairing channels. With the fermionic representation $\mbf S_{i\tau} = \sum_{\s\til\s}c_{i\tau\s}^\dag(\bsb\s_{\s\til\s}/2)c_{i\tau\til\s}$, a four fermion interaction is obtained
\beeq
H_J = -\fr{J}{4}\sum_i\sum_{\tau\tau'}\sum_{\s\til\s\s'\til\s'}\lf(\bsb\s_{\s\til\s}\cdot\bsb\s_{\s'\til\s'}\ri)c_{i\tau\s}^\dag c_{i\tau'\s'}^\dag c_{i\tau'\til\s'}c_{i\tau\til\s}.
\eneq
The identity $\bsb\s_{\s\til\s}\cdot\bsb\s_{\s'\til\s'}=2\d_{\s\til\s'}\d_{\s'\til\s}-\d_{\s\til\s}\d_{\s'\til\s'}$ is exploited to decompose the Pauli matrices. While the second term enhances the onsite repulsion $U$ trivially, the first term provides a nontrivial perturbation.

We ignore the trivial part and apply the Fierz identity. While the orbital representation exhibits the normal inner product $\vec\tau(i\tau^y)\cdot(i\tau^y)^\dag\vec\tau^\dag$, the spin representation is transposed $[\vec\s(i\s^y)]^T\cdot(i\s^y)^\dag\vec\s^\dag$. An expression in terms of pairing channels arises
\beeq
H_J = \fr{J}{16}\sumvdp{k}
\lf[-(\vec P_{\mbf k}^{\tau^0\bsb\s})^\dag\cdot\vec P_{\mbf k'}^{\tau^0\bsb\s}+(\vec P_{\mbf k}^{\bsb\tau\s^0})^\dag\cdot\vec P_{\mbf k'}^{\bsb\tau\s^0}\ri],
\eneq
where the antisymmetric pairing is demanded due to Fermi statistics. Notice a splitting of onsite repulsion $\d U\sim\pm J$ between the spin singlet and triplet pairings. The spin singlet pairing channels gain larger onsite repulsion $U$, thereby experiencing a stronger attractive interaction after the Kohn-Luttinger renormalization. We conclude that the $d$-wave superconductivity near Van Hove filling manifests the spin singlet pairing.

The preference of spin singlet pairing due to Hund's coupling deserves a special discussion. In the bare theory with Hund's coupling, the spin triplet pairings gain lower energy than the spin singlet pairing. A dominant $d$-wave superconductivity with spin triplet and orbital singlet pairings is therefore identified in an effective model with strong coupling \cite{xu18prl}. However, the experimental results indicate a spin singlet pairing in the superconductivity of twisted bilayer graphene \cite{cao18nsc}. This can be demonstrated in models with either a violation of Hund's first rule \cite{dodaro18prb} or an anti-Hund's coupling \cite{,you18ax}. We argue that the anti-Hund's coupling can be regarded as a result of Kohn-Luttinger renormalization. In the pairing channels with superconductivity, the splitting of degeneracy between spin singlet and triplet pairings is determined by the second order corrections Eq.~(\ref{eq:intpairchan}). The bare Hund's coupling Eq.~(\ref{eq:hunds}) is converted to an effective anti-Hund's coupling $\til J\sim -UJ$ under Kohn-Luttinger renormalization. With this effective anti-Hund's coupling, the spin singlet pairing is favored in the superconductivity. Notice that this effect does not occur in the normal channels, where the interaction remains repulsive.

The degeneracy can be further reduced by the perturbative pair hopping interaction \cite{yuan18prb}
\beeq
H_{J'}
= \fr{J'}{2}\sum_{i,\tau\tau'\s\s'}c_{i\tau\s}^\dag c_{i\tau\s'}^\dag c_{i\tau'\s'}c_{i\tau'\s}
\eneq
with $J'\ll U$. While the spin representation remains the original inner product, the orbital representation exhibits a single product $[(-i\tau^y)(i\tau^y)][(i\tau^y)^\dag(-i\tau^y)^\dag]$. The expression
\beeq
H_{J'} = \fr{J'}{8}\sumvdp{k}(P_{\mbf k}^{\tau^y\s^0})^\dag P_{\mbf k'}^{\tau^y\s^0}
\eneq
indicates the dominance of a single pairing channel. With $\tau^y=(i/\sqrt2)[(-\sqrt2\tau^+)+\sqrt2\tau^-]$, we identify the orbital pairing as $\ket{1\,0}_y=(i/\sqrt2)(\ket{1\,1}+\ket{1\,-1})$. Such perturbation from the pair hopping interaction can also be identified as a part of the effective anti-Hund's coupling.

We have seen a breakdown of degeneracy due to the introduction of bare Hund's coupling. After the Kohn-Luttinger renormalization, the bare Hund's coupling is converted into an effective anti-Hund's coupling, thereby favors the spin singlet pairing. Despite the original six-fold degeneracy in the quantum number pairings, only one pairing channel dominates under the perturbations, and this channel is spin singlet.

\subsubsection{Ginzburg-Landau theory}

There is still a two-fold degeneracy due to different $d$-wave configurations. The breakdown of this degeneracy can be analyzed through the Ginzburg-Landau theory \cite{nandkishore12np}. Writing the partition function as a coherent path integral and applying the Hubbard-Stratonovich transformation, we derive the free energy near the critical temperature $T_c$ (see Appendix \ref{sec:appgl})
\beeq
\beal
F[\bar\D_{1,2},\D_{1,2}]
&= r\lf(|\D_1|^2+|\D_2|^2\ri)
+u\lf(|\D_1|^2+|\D_2|^2\ri)^2
\\&\quad
-\fr{u}{3}\lf[2|\D_1|^2|\D_2|^2-\lf(\D_1^2\bar\D_2^2+\bar\D_1^2\D_2^2\ri)\ri].
\enal
\eneq
Each order parameter $\D_i$ corresponds to a $d$-wave pairing channel. While the quadratic coefficient $r=\a(T-T_c)$ with $\a>0$ changes sign across the critical temperature $T_c$, the quartic coefficient $u$ is always positive. The minimal free energy occurs when $|\D_1|=|\D_2|=\D$ and $\D_2/\D_1=\pm i$. Correspondingly, the gap function in the patch representation is
\beeq
\D^d_{\eta\eta'} = (\tau^y\s^0)_{\eta\eta'}\D\fr{1}{\sqrt3}(1,e^{\pm2\pi i/3},e^{\mp2\pi i/3}).
\eneq
The phase of the gap function exhibits a winding $e^{\pm2i\phi}$ along the Fermi surface, where $\phi$ is the polar angle in the momentum space. A phase $\pm4\pi$ is gained after a full winding. This state can be identified as a $d+id$ chiral superconductivity with broken time reversal symmetry and nontrivial topological features \cite{nandkishore12np, xu18prl}.

\subsection{$f$-wave superconductivity}

The $f$-wave pairing is symmetric in quantum numbers (therefore antisymmetric in its momentum space structure), and so its degeneracy cannot be lifted by onsite perturbations. To break the degeneracy, we must introduce extended interactions. Consider the extended Hubbard model with significantly decaying repulsions. Previous investigation \cite{nandkishore14prb} indicates a proportionality of leading correction to the second neighbor repulsion $\d g^f\sim-U_2$. Introduce a spin exchange interaction
\beeq
H_{J_2} = J_2\sum_{\epvl{\epvl{ij}}}\sum_{\tau\tau'}\mbf S_{i\tau}\cdot\mbf S_{j\tau'}
\eneq
with $J_2\ll U_2$. By similar decomposition to the treatment of Hund's coupling, we arrive at the expression
\beeq
H_{J_2} = \fr{J_{2,\mbf k-\mbf k'}}{16}\sumvdp{k}
\lf[-(\vec P_{\mbf k}^{\vec\tau\s^0})^\dag\cdot\vec P_{\mbf k'}^{\vec\tau\s^0}+(\vec P_{\mbf k}^{\vec\tau\bsb\s})^\dag\cdot\vec P_{\mbf k'}^{\vec\tau\bsb\s}\ri]
\eneq
and an additional trivial correction $-J_2$ to the repulsion $U_2$. For most materials described by the Hubbard models, the spin exchange exhibits the antiferromagnetic feature $J_2>0$. The repulsion $U_2$ is suppressed more in the spin singlet pairing channel. Therefore, the $f$-wave superconductivity manifests the spin singlet pairing. When the opposite situation occurs, the spin triplet pairing is favored by a ferromagnetic spin exchange $J_2<0$. Notice that the superconductivity is triggered only when the Kohn-Luttinger renormalization overcomes the bare extended repulsion.

\section{Discussion and conclusion}
\label{sec:concl}

We have demonstrated how superconductivity emerges from weak electronic repulsion in an effective two orbital honeycomb superlattice model.  The pairing channels exhibit an approximate $\mrm{SU(4)}$ symmetry. Utilizing the patch models, we conduct the Kohn-Luttinger renormalization to probe the potential superconductivities in a broad range of fillings. Near the Van Hove filling, the dominant pairing channel is $d$-wave, whereas away from Van Hove filling it is $f$-wave. We have investigated the lifting of degeneracy by perturbations. We have shown that a bare Hund's coupling is converted to an effective anti-Hund's coupling under Kohn-Luttinger renormalization, and thereby selects a single spin singlet pairing channel for the $d$-wave pairing. Performing also a Ginzburg-Landau analysis, we therefore predict a spin singlet $d+id$ pairing state close to Van Hove filling.  Away from Van Hove filling, $f$-wave pairing dominates. The degeneracy now is only lifted when extended interactions are taken into account. With an antiferromagnetic spin exchange, spin singlet pairing is favored.

Our study provides a clean and systematic analysis of superconductivity born of weak repulsion in an effective model for twisted bilayer graphene. How robust the conclusions are to the details of the model remains to be established, but it may be hoped that the conclusions are robust. One important feature that has {\it not} been addressed is the insulating state observed near the superconducting dome \cite{cao18nmi,cao18nsc}. This insulating state may be related to the density wave states in the weak coupling regime. Exploring the competition between superconducting and density wave states is an important open problem, that we however leave to future work \cite{lin18tbp}.

\section{Acknowledgements}
This research was sponsored by the Army Research Office and was accomplished under Grant No. W911NF-17-1-0482. The views and conclusions contained in this document are those of the authors and should not be interpreted as representing the official policies, either expressed or implied, of the Army Research Office or the U.S. Government. The U.S. Government is authorized to reproduce and distribute reprints for Government purposes notwithstanding any copyright notation herein.

\appendix

\section{Kohn-Luttinger renormalization}
\label{sec:appkl}

In the Kohn-Luttinger renormalization, the static part $\Pi_{\mbf q}$ of the polarization bubble
\beeq
\Pi_{\mbf q\til\o} = -T\sum_\o\intv{p}G_{\mbf p\o}G_{(\mbf p+\mbf q)(\o+\til\o)}
\eneq
is defined by the limit $\til\o\rar0$. Here $G_{\mbf p\o}=1/(i\o-\xi_{\mbf p})$ is the free fermionic propagator. The Matsubara frequencies $\o$ and $\til\o$ correspond to the fermionic and bosonic modes, respectively. After the Matsubara frequency summation, the polarization bubble is transformed into
\beeq
\Pi_{\mbf q\til\o} = \intv{p}\fr{f(\xi_{\mbf p+\mbf q})-f(\xi_{\mbf p})}{i\til\o-(\xi_{\mbf p+\mbf q}-\xi_{\mbf p})},
\eneq
where $f(z)=1/(e^{z/T}+1)$ is the Fermi function.

We follow the analysis of different pairing channels at various fillings in Ref.~\onlinecite{nandkishore14prb}. Away from Van Hove filling, the Fermi surface is not nested, and the density of states is finite. Since all patch pairs provide similar contributions, the corrections can be analyzed by counting the numbers $N_{\mbf q}$ of patch pairs involved in the polarization bubbles. Things become significantly different in the vicinity of Van Hove filling. Whenever a patch pair experiences the divergent density of states, a logarithmically divergent factor $\ln(\L/T)$ arises. Here the ultraviolet cutoff $\L$ corresponds to the size of patches. The same factor also shows up when the patch pairs access the nesting of Fermi surface, where $\xi_{\mbf p+\mbf q}=-\xi_{\mbf p}$ for momenta $\mbf p$'s in the patches. Due to these divergent characteristics, the corrections become remarkably different and requires special studies.

\subsubsection{Near half filling}

When the system is close to the half filling $\mu\gtrsim0$, the patches are approximated by the six Dirac points. In this approximation, the momenta $\mbf Q$ and $\mbf Q+2\mbf K$ are equivalent to $\mbf 0$ and $2\mbf K$ up to some reciprocal lattice vectors. The corresponding polarization bubbles can also be identified with each other, where $\Pi_{\mbf 0}=\Pi_{\mbf Q}$ and $\Pi_{2\mbf K}=\Pi_{\mbf Q+2\mbf K}$. While the zero momentum $\mbf0$ links $N_{\mbf0}=6$ patches to themselves, each momentum $2\mbf K$ connects $N_{2\mbf K}=3$ pairs of patches. This counting implies an approximate relation between the polarization bubbles $\Pi_{2\mbf K}\apx\Pi_{\mbf 0}/2$. From Eq.~(\ref{eq:intpairchan}), we find attractive corrections to interaction in both $s$-wave and $f$-wave pairing channels. The $s$-wave pairing channel remains repulsive due to the bare repulsion, while the $f$-wave pairing channel exhibits a nontrivial attractive interaction
\beeq
g^{f} \apx \fr{9U^2}{16}\Pi_{\mbf 0}.
\eneq
Therefore, the $f$-wave superconductivity dominates in this regime. Notice that the polarization bubble $\Pi_{\mbf 0}$ is determined by the density of states. The proportionality to the chemical potential $\Pi_{\mbf 0}\sim\mu$ indicates the onset of superconductivity only at finite filling $\mu>0$.

\subsubsection{Near Van Hove filling}

The features of superconductivity vary significantly when the system is close to Van Hove filling $\mu=t$. We first examine the corrections in patch model II. As the system is finitely distant from Van Hove filling, the counting of patch pairs still applies. With the numbers of patch pairs for linking momenta $N_{\mbf 0}=6$, $N_{2\mbf K}=1$, and $N_{\mbf Q}=N_{\mbf Q+2\mbf K}=2$, the retention of $f$-wave superconductivity is confirmed. Notice the potential degeneracy with $p$-wave pairing channel in these regimes. When the extended repulsions are introduced, the dominance of $f$-wave pairing channel is retrieved \cite{nandkishore14prb}. The corrections in the vicinity of Van Hove filling require special studies. Due to the nesting of Fermi surface, the polarization bubble $\Pi_{2\mbf K}$ becomes logarithmically divergent. The interaction in the $f$-wave pairing channel becomes negative, and the corresponding superconductivity disappears. However, attractive corrections with logarithmic divergences $\ln(\L/T)$ arise in the $s$-wave and $d$-wave pairing channels. While the $s$-wave pairing channel remains repulsive due to the bare repulsion, the $d$-wave pairing channel acquires an attractive interaction. Therefore, the $d$-wave superconductivity can arise in this regime.

The examination of patch model I is also necessary. When the filling is away from Van Hove filling, the polarization bubble $\Pi_{\mbf 0}$ is the largest, and the $f$-wave superconductivity is dominant \cite{nandkishore14prb}.  When the Van Hove filling is approached, the divergent density of states results in logarithmic divergences $\ln(\L/T)$ in the polarization bubbles $\Pi_{\mbf 0}=\Pi_{2\mbf K}$. The other two polarization bubbles $\Pi_{\mbf Q}=\Pi_{\mbf Q+2\mbf K}$ gain the additional access to the nesting of Fermi surface, thereby manifest the more divergent scaling $\ln^2(\L/T)$. The only attractive correction with divergence $\ln^2(\L/T)$ arises in the $s$-wave pairing channel. However, the second order correction can not exceed bare interaction in perturbation theory. Therefore, there is no superconductivity in patch model I, and the only pairing instability arises in the $d$-wave channel from the states captured by patch model II.

\section{Ginzburg-Landau theory}
\label{sec:appgl}

In this section, we derive the free energy in terms of order parameters near the critical temperature $T_c$. The analysis starts with the coherent path integral formulation of partition function
\beeq
Z = \int\mca D(\psi^\dag,\psi)e^{-S[\psi^\dag,\psi]},
\eneq
where the action is
\beeq
\beal
S[\psi^\dag,\psi]
= \int_\tau&\Bigg\{
\sum_{\k=\pm}\psi_\k^\dag(\p_\tau+\xi_\k)\psi_\k
\\
&-\fr{1}{2}g^d\lf[(P^{d_1})^\dag P^{d_1}+(P^{d_2})^\dag P^{d_2}\ri]
\Bigg\}.
\enal
\eneq
With the Hubbard-Stratonovich transformation, the quartic interaction is decoupled by the bosonic order parameters $\D_1$ and $\D_2$. Define the Nambu spinor
\beeq
\Psi = \lf(\bear{c}\psi_+\\ \g(\psi_-^\dag)^T \enar\ri)
\eneq
and the inverse Gor'kov Green's function
\beeq
\mca G^{-1} = \lf(\bear{cc}-\p_\tau-\xi_+&\sum_i\D_id^{(i)}\tau^y\s^0\\\sum_i\bar\D_id^{(i)}\tau^y\s^0&-\p_\tau+\xi_-\enar\ri).
\eneq
The partition function is expressed as a path integral of Nambu spinor and order parameter
\beeq
Z = \int\mca D(\bar\D_{1,2},\D_{1,2})\mca D(\Psi^\dag,\Psi)e^{-S[\bar\D_{1,2},\D_{1,2},\Psi^\dag,\Psi]},
\eneq
where the action becomes
\beeq
S[\bar\D_{1,2},\D_{1,2},\Psi^\dag,\Psi]
= \int_\tau\lf(\fr{2}{g^d}\sum_i|\D_i|^2-\Psi^\dag\mca G^{-1}\Psi\ri).
\eneq
Impose the static condition $\D_i(\tau)=\D_i$ and convert to the Matsubara frequency representation $\Psi(\tau)=\sqrt T\sum_\o\Psi_\o e^{-i\o\tau}$. Integrating out the Nambu spinor, we arrive at the partition function
\beeq
Z = \int\mca D(\bar\D_{1,2},\D_{1,2})e^{-F[\bar\D_{1,2},\D_{1,2}]/T}
\eneq
along with the free energy
\beeq
F[\bar\D_{1,2},\D_{1,2}]
= \fr{2}{g^d}\lf(|\D_1|^2+|\D_2|^2\ri)-\Tr\ln\mca G^{-1}.
\eneq
Notice that the identity $\ln\det\mca G^{-1}=\Tr\ln\mca G^{-1}$ has been utilized. The Gor'kov Green's function in momentum frequency space representation is
\beeq
\mca G^{-1} = \lf(\bear{cc}G_+^{-1}&\sum_i\D_id^{(i)}\tau^y\s^0\\\sum_i\bar\D_id^{(i)}\tau^y\s^0&G_-^{-1}\enar\ri),
\eneq
where we define the free electron and hole propagators as $G_\pm=1/(i\o\mp\xi_\pm)$.

We expand the free energy in the vicinity of critical temperature $T_c$. Define $\mca G_0^{-1}=\mca G^{-1}(\D_{1,2}=0)$ and $\hat\D=\mca G^{-1}-\mca G_0^{-1}$. Ignoring the constant part, the expansion up to quartic order takes the form
\beeq
\beal
F[\bar\D_{1,2},\D_{1,2}]
&= \fr{2}{g^d}\lf(|\D_1|^2+|\D_2|^2\ri)
\\&\quad
+\fr{1}{2}\Tr(\mca G_0\hat\D)^2+\fr{1}{4}\Tr(\mca G_0\hat\D)^4,
\enal
\eneq
where
\beeq
(\mca G_0\hat\D)^2=\sum_{ij}d^{(i)}d^{(j)}\mrm{diag(G_+G_-\D_i\bar\D_j,G_-G_+\bar\D_i\D_j)}
\eneq
serves as the small parameter of the expansion. With the nonzero quartic traces $\Tr(d^{(1)})^4=\Tr(d^{(2)})^4=1/2$ and $\Tr[(d^{(1)})^2(d^{(2)})^2]=\Tr(d^{(1)}d^{(2)}d^{(1)}d^{(2)})=1/6$, we arrive at the free energy
\beeq
\beal
F[\bar\D_{1,2},\D_{1,2}]
&= r\lf(|\D_1|^2+|\D_2|^2\ri)
+u\bigg[\lf(|\D_1|^4+|\D_2|^4\ri)
\\&\quad
+\fr{4}{3}|\D_1|^2|\D_2|^2+\fr{1}{3}\lf(\D_1^2\bar\D_2^2+\bar\D_1^2\D_2^2\ri)
\bigg].
\enal
\eneq
The quadratic coefficient $r=2/g^d+4\Tr(G_+G_-)$ exhibits a linear scaling $r=\a(T-T_c)$ with $\a>0$, and a change of sign occurs at the critical temperature $T_c$. Meanwhile, the quartic coefficient $u=\Tr(G_+G_-G_+G_-)$ is always positive so that the theory is stable.

The symmetry breaking in the ordered phase $T<T_c$ is derived from the minimization of free energy. Denote the relative phase between the two order parameters as $\mrm{Arg}(\D_2/\D_1)=\t$. The free energy can be rewritten as
\beeq
\beal
F[\bar\D_{1,2},\D_{1,2}]
&= u\lf(|\D_1|^2+|\D_2|^2+\fr{r}{2u}\ri)^2-\fr{r^2}{4u}
\\&\quad
-\fr{4u}{3}\sin^2\t|\D_1|^2|\D_2|^2,
\enal
\eneq
and the minimum occurs at $|\D_1|=|\D_2|$ with $\t=\pm\pi/2$.

\bibliography{Reference}

\end{document}